# Malicious User Detection in Spectrum Sensing for WRAN Using Different Outliers Detection Techniques

Manish B Dave[#1], Mitesh B Nakrani[#2]

[1]*Assistant Professor, C. U. Shah College of Engg. & Tech., Wadhwan city-363030, Surendranagar Gujarat India*

[2]*Assistant Professor, C. U. Shah College of Engg. & Tech., Wadhwan city-363030, Surendranagar Gujarat India*

*Abstract—* **In cognitive radio it is of prime importance that the presence of Primary Users (PU) is detected correctly at each of the time. In order to do so the help from all present Secondary Users (SU) is taken and such a taken is known as co-operative spectrum sensing. Ideally it is assumed that all the secondary users give the correct result to the control center. But there are certain conditions under which the secondary users deliberately forward wrong result to the control center so as to degrade the performance of the cognitive network. In this paper we study the different techniques for detecting the malicious users or outliers. We take into consideration practical environmental condition such that the received signal of the secondary users is made to undergo fading and noise is also introduced in the signal. We further go on to examine each of the outlier detector techniques and find out the most suitable at various instants.**

*Keywords*- Cognitive Radio, Co-operative Spectrum Sensing, WRAN, Cyclo-stationary detector, Malicious Users, Outliers, Median Absolute Deviation (MAD), Medcouple, Skewness.

## I. INTRODUCTION

Cognitive Radio is a new technology which is still in its early stages. It is technology wherein the licensed bands are used by the unlicensed users thereby better utilizing the underutilized bands [1]. In cognitive radio, a frequency band is checked for the presence of primary users (PU) and if empty the unlicensed users or the secondary users (SU) are allowed to operate in the band. The detection poses a serious performance issue as each of the PU's use different modulation and transmitted power. There are number of techniques for detecting the presence of PU in the frequency band which are elaborated in [2]. In order to improve the performance of the system results regarding the presence of PU are not limited to a single sensing SU but rather at different sensing SUs which is known as co-operative spectrum sensing. In this method the sensed data from different points is brought to a control centre wherein the final decision is made. A significant performance gain is achieved when using this method [3-4]

In ideal case the information from various SU's is made available to the control centre and final decision is made by averaging all the information. But in practical conditions there are many such instances when a SU deliberately reports false information to the control centre [5]. Such users are known as malicious users. As a result of this false report the performance of the system is degraded substantially. In statistics generally data which are far away from the mean distribution are classified as outliers. In the case of malicious users problems they can be classified as outlier because they will give information which will vary in large amount from the mean distribution of the accumulated data. Thus detecting these malicious users can be considered as simply detecting outliers in a given distribution. There number of methods for detecting outliers which are elaborated in [6-7].

The paper is organized in 4 sections. Following this introduction section a number of outlier detection methods are discussed in section II, which are used to elaborate the performance in terms of detecting the malicious users. System model used to find out the results is discussed in section III. The performance of different outlier detection methods are presented in section IV. This is followed by conclusion in section IV.

The paper proposes to find the outlier detection method which gives better performance in comparison to others under different number of malicious users.

## II. OUTLIER DETECTION TECHNIQUES

A node can be malicious due to device fault or for selfish reasons. In case of selfish reasons it might happen that a SU reports that a PU is present and force the control centre to such a decision and use the channel to transmit its own signal. Such types of malicious users are known as "Always Yes" users. On the other hand in order to cause interference to the PU some SU's might force the control centre to think that the channel is empty. Such users can be classified as "Always No" users [7-8]. Another subcategory of malicious users can be malicious at specific instants of observation and can be either "Yes or No".

In case of "Always Yes" users the reported energy level will always be much greater than the threshold level decided by the control centre and will deviate by a large amount from levels of rest users. In case of "Always No" users the reported energy will be much lesser than that of the threshold. According to the definitions of outliers from [9] "An outlier is an observation which deviates so much from the other observations as to arouse suspicions that it was generated by different mechanisms". So the levels reported by malicious users can be classified as outliers.

There are number of methods to detect the outliers some of which are elaborated further in this section and are used to remove the malicious users from being considered for the final decision.

*1) Mean Difference (MD):*

Variance are good measure for checking the variability in a set of data but it gives superior results in scope to the data which are almost or exactly normally distributed. Mean difference doesn't have such limitation [10]. The mean difference is given as

$$MD = E\{|(X1 - X2)|\}$$

where X1 and X2 being data at two different instant.

For the case of random sample of size 'N', the mean difference can be calculated as the arithmetic mean of the absolute difference of all values as

$$MD = \frac{1}{N^2} \sum_{i=1}^{N} \sum_{j=1}^{N} |x_i - x_j|$$

*2) Medain Absolute Deviation (MAD):*

MAD is similar to standard deviation in terms of estimating the spread of data in a given distribution but has a breakdown point of 50% like median. It gives the median of absolute deviation of data values from the median of the sample under consideration. MAD is given as [11]

$$MAD = median|x_i - median(x_i)|$$

*3) Sn Estimator (Sn):*

It has been found that MAD gives better result when the sample distribution is symmetrical. Thus this property can be of significant disadvantage in case of skewed distribution. So for skewed distributions, better estimators are proposed in [12]. First of them is the Sn estimator which is given as

$$S_n = 1.192 * Median_i \left[ Median_j |x_i - x_j| \right]$$

Sn is similar to MAD in manner that both use combination of medians. In terms of difference, MAD takes into account the deviation from the central value of the distribution whereas Sn estimator considers the distances between any two observations.

*4) Qn Estimator (Qn):*

It is similar to the Sn estimator and has the same properties while the only exception that it smooth's out any discontinuities in the result obtained from Sn estimator. The Qn estimator is given as

$$Q_n = 2.2 * \left( 1^{st} Quartile |x_i - x_j| \right) : i < j$$

*5) Medcouple (MC):*

In case of malicious users false notification the reported level distribution becomes skewed. In order to measure this skewness we can apply the medcouple [13]. The medcouple (MC) is defined as

$$MC(S) = \underset{xi<mS<xj}{Median} \left[ \frac{(x_j - mS) - (mS - x_i)}{x_j - x_i} \right]$$

$$Q_1 - h_l(MC)IQR; Q_3 + h_r(MC)IQR$$

where $x_i$ and $x_j$ are sampled independently from the distribution S, mS is the median of the distribution and IQR is the inter-quartile range.

It can be said from the definition that the value of medcouple will lie between -1 and +1. For a right skewed distribution it will have a positive value and for left skewed the value will be negative.

For the present case we take into account the exponential model of the medcouple which is given as

$$h_l = 1.5 * \exp(-3.5 * MC)$$
$$h_r = 1.5 * \exp(4 * MC)$$

Note that in each of the schemes the $x_i$ and $x_j$ are reported energy levels of different SU's under consideration.

### III. SYSTEM MODEL

We consider a group of N secondary or cognitive users. Practical environment are taken into consideration by taking into account the effect of noise and fading. All the SU's measure the energy of the channel at their respective sites and forward their measured energy level to the control or the decision centre via control channels. It has been assumed that control channels are perfect. At the control centre the different measurement data are collected and data fusion methods are used to obtain the final decision.

Since main aim of cognitive radio is to provide minimal interference to the PU's we have considered only "Always No" users in this paper. We have assumed that the malicious users produce extreme false values. Continuing from our previous work SU's measure the channel energy with the help of cyclo-stationary detector work [14]. The result is sent to the control centre and averaging scheme is used to obtain the detection probability. Based on the threshold obtained from the different outlier detection methods malicious users are detected and they are ignored when making the final decision at the control center.

## IV. SIMULATION RESULTS

We consider a system having 50 Secondary Users (N=50). The average received SNR at each of the SU location is -10 dB. In addition Rayleigh fading is introduced between PU transmitter and SU's. The energy levels are obtained with the help of the cyclo-stationary detector [14]. Now to the energy level of all the SU's each of the outliers detection methods are applied and a threshold is decided for each of the techniques. The threshold obtained is compared with the energy level of each of the SU's. If the reported level is less than threshold then it is classified as outlier or malicious user.

Different time instants are considered such that the fading coefficient for each of the N users varies and threshold is calculated for each of the sampling instant. Figure 1 displays the different values of threshold for 50 different time instants with number of malicious users equal to '10'.

The iterative procedure is continued till a stopping condition is reached. The condition can be maximum number of iteration or difference between two thresholds values be less than a predefined tolerance.

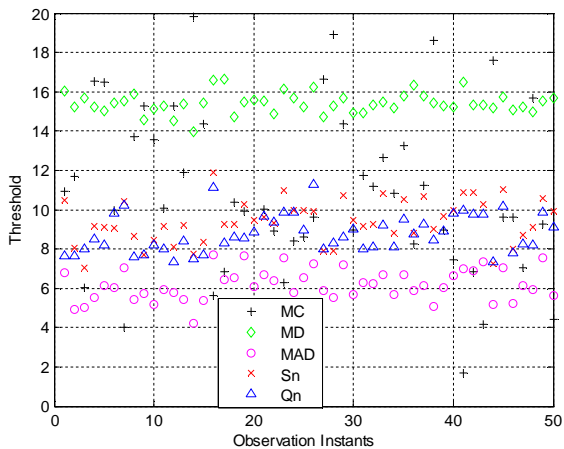

Fig. 1 Threshold values for different outliers methods for 50 observations with 10 malicious users.

We considered 50 different observation instants and counted the number of malicious users detected by a particular outlier method each time. In figure 2 plot of the number of malicious users versus the number of times the outlier techniques have detected that number of malicious users.

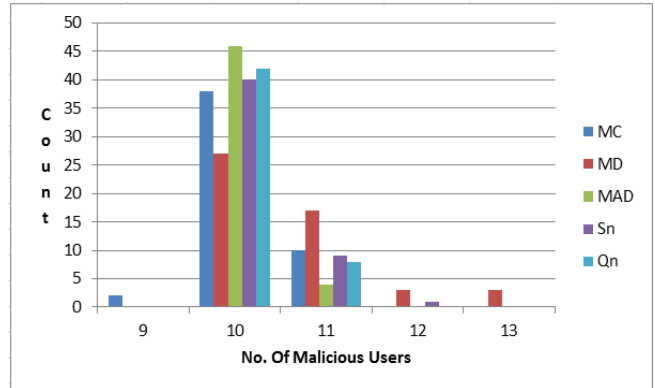

Fig. 2 No of times correct detection of malicious users for different outliers techniques (actual outliers=10).

Figure 3 displays the threshold levels at different instants with number of malicious users equal to 5. For the threshold levels obtained the count for 50 observations is plotted in figure 4.

We have considered different number of malicious users and found out how many times different techniques correctly identify the malicious users. Figure 5 displays the case when the actual number of malicious users is equal to 7 whereas the situation of actual number of malicious users is depicted by figure 6.

Table 1 shows the correct detection rates for various outlier detection techniques. It presents good comparison between different methods as to which performs better in terms of malicious user detection. It can be inferred from the table that the most consistent and accurate of the five techniques is the Median Absolute Deviation or the MAD. This is followed by Sn and Qn estimator as there is barely and difference between the two. Next in the list is the Mean Difference (MD) as for less number of times it has detected correctly as compared to MAD followed by medcouple.

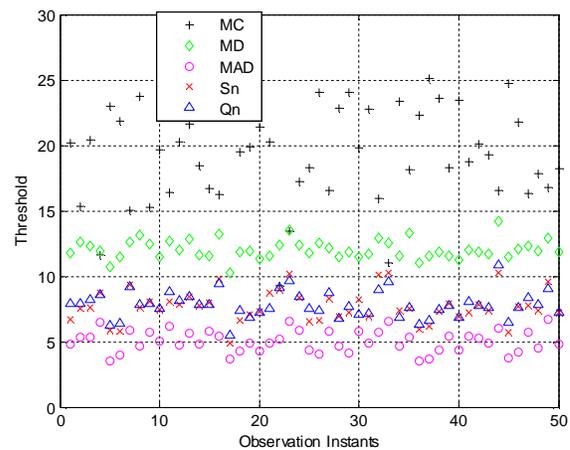

Fig. 3 Threshold values for different outliers methods for 50 observations with 5 malicious users.

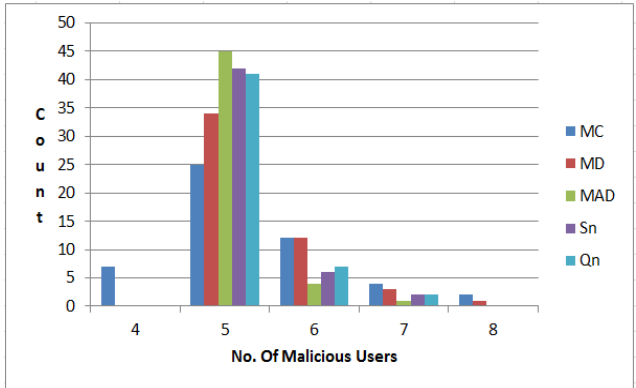

Fig. 4 No of times correct detection of malicious users for different outliers techniques (actual outliers=5).

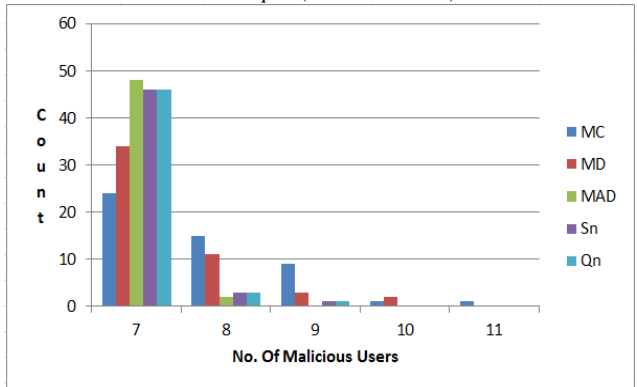

Fig. 5 No of times correct detection of malicious users for different outliers techniques (actual outliers=7).

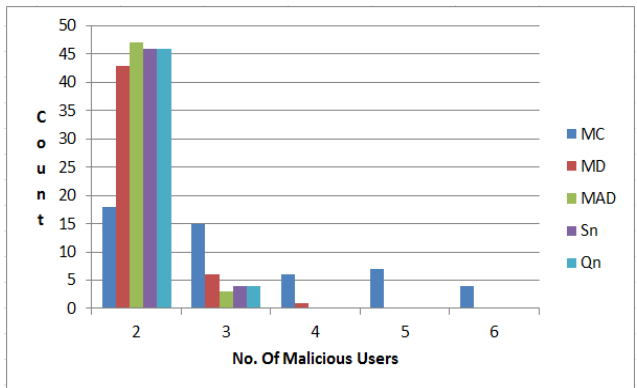

Fig. 6 No of times correct detection of malicious users for different outliers techniques (actual outliers=2).

Figure 7 gives the plot of detection probability versus the threshold (ROC) for malicious users equal to '5'. The black dash dot (WM) plots the detection probability while considering the malicious users. All the detection schemes give better result than in the presence of malicious users. The medcouple gives the highest probability because half of the users have been considered malicious. The most consistent scheme MAD gives better result than WM but no as high as MC. In figure 8 number of malicious users considered is '10'. Since in this case the worst performing scheme is Mean Difference, the detection is high as almost half of users have been considered malicious.

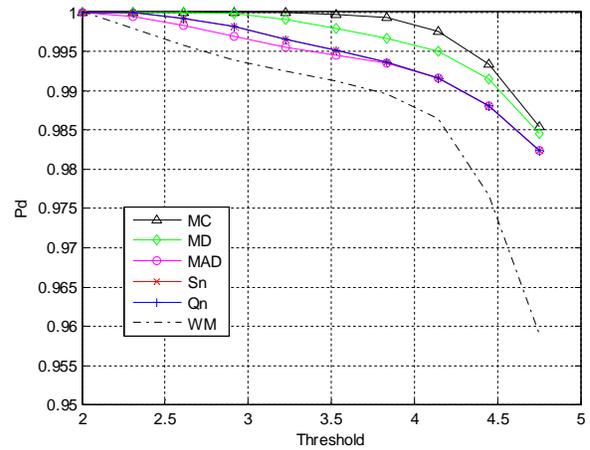

Fig. 7 Pd vs Th (ROC) for malicious users =5

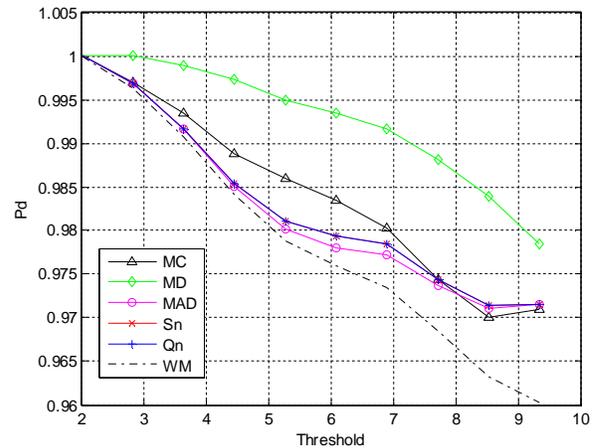

Fig. 8 Pd vs Th (ROC) for malicious users =10

## V. CONCLUSION

The simulation results clearly indicate that the ROC curve cannot be the only plot upon which the performance of an outlier detection scheme can be judged. The reason being that in both cases the maximum probability of detection is for those schemes in which the number of malicious users detected is much greater than actual value. Thus for checking the performance of the different outlier techniques

TABLE I
CORRECT DETECTION BY VARIOUS SCHEMES

| Actual No. of Malicious Users | Number of times correct detection of malicious users (out of 50) | | | | |
|---|---|---|---|---|---|
| | (MC) | (MD) | (MAD) | (Sn) | (Qn) |
| 2 | 18 | 43 | 47 | 46 | 46 |
| 5 | 25 | 34 | 45 | 42 | 41 |
| 7 | 24 | 34 | 48 | 46 | 46 |
| 10 | 38 | 27 | 46 | 40 | 42 |

the number of times correctly detected malicious users must also be considered. From the results we can conclude that the MAD scheme is good at detecting the exact number of malicious users and also has a good detection rate

REFERENCES

[1] FCC, "Spectrum Policy Task Force Report," vol. ET Docket No. 02-155, November 2002.

[2] T. Yucek and H. Arslan, "A survey of spectrum sensing algorithms for cognitive radio applications," *Communications Surveys Tutorials, IEEE*, vol. 11, no. 1, pp. 116-130, quarter 2009..

[3] S. Atapattu, C. Tellambura, and Hai Jiang, "Energy Detection Based Cooperative Spectrum Sensing in Cognitive Radio Networks," *Wireless Communications, IEEE Transactions on*, vol. 10, no. 4, pp. 1232-1241, april 2011.

[4] A. Ghasemi and E.S. Sousa, "Collaborative spectrum sensing for opportunistic access in fading environments," , nov. 2005, pp. 131-136.

[5] S. M. Mishra, A. Sahai and R. W. Brodersen, "Cooperative sensing among cognitive radios," *IEEE International Conference on Communication ICC'06"*, vol. 4, pp. 1658-1663, June 2006.

[6] R. Chen, J.-M. Park, and J. Reed, "Defense against PU emulation attacks in CR networks," *IEEE J. Sel. Areas Commun.*, vol. 26, pp. 2537, Jan. 2008.

[7] P. Kaligineedi, M. Khabbazian and V. K. Bhargava, "Secure Cooperative Sensing Techniques for Cognitive Radio System" *IEEE International Conference on Communications*, Beijing, 19-23 May2008, pp. 3406-3410.

[8] Kaligineedi, P., M. Khabbazian, and V. K. Bhargava, "Malicious User Detection in a Cognitive Radio Cooperative Sensing System", *Wireless Communications, IEEE Transactions on*, vol. 9, no. 8, pp. 2488 -2497, aug., 2010.

[9] HAWKINS,D.M. *Identification of Outliers*. Chapman and Hall, London and New York,1980.

[10] Yitzhaki, Shlomo, Gini's Mean difference: a superior measure of variability for non-normal distributions, *Metron - International Journal of Statistics*, LXI, issue 2, p. 285-316, 2003.

[11] Seo Songwon, "A Review and Comparison of Methods for Detecting Outliers in Univariate Data Sets", Master's Thesis, University of Pittsburgh, 2006.

[12] PJ Rousseeuw, C Croux, "Alternatives to the median absolute deviation", *Journal of the American Statistical Association*, pp. 1273-1283. DOI:10.1080/01621459.1993.10476408

[13] Vanderviere, E. & Huber, M. An adjusted boxplot for skewed distributions. *COMPSTAT2004 Symposium: proceedings in computational statistics*(pp. 1933-1940), 2004.

[14] M. B. Dave, "Optimal number of users in Co-operative spectrum sensing in WRAN using Cyclo-Stationary Detector", *International Journal of Engineering Trends and Technology(IJETT)*, V4(7), pp. 2806-2810 July 2013. ISSN:2231-5381. arXiv:1403.3312v1[cs.NI].